\documentclass[twocolumn,aps,prl,nofootinbib,superscriptaddress]{revtex4-2}
\usepackage{amsmath}
\usepackage{bm}
\usepackage{graphicx}
\usepackage{mathrsfs}
\usepackage{multirow}
\usepackage[normalem]{ulem}
\usepackage{graphicx}
\usepackage{footnote}
\usepackage{booktabs, ctable}
\usepackage{xcolor, framed}
\usepackage[colorlinks=true,
            linkcolor=red,
            citecolor=blue,
            urlcolor=black]{hyperref}

\graphicspath{{./figures/}}

\begin{document}

\title{
Accessing Exotic Hadronic States via Charmed-Meson Femtoscopy in Relativistic Heavy-Ion Collisions}

\author{Jiaxing Zhao}
\email{jzhao@itp.uni-frankfurt.de}
\affiliation{Helmholtz Research Academy Hesse for FAIR (HFHF), GSI Helmholtz Center for Heavy Ion Physics, Campus Frankfurt, 60438 Frankfurt, Germany}
 \affiliation{Institut f\"{u}r Theoretische Physik, Johann Wolfgang Goethe-Universität,Max-von-Laue-Straße 1, D-60438 Frankfurt am Main, Germany}

\author{Taesoo Song}
\email{t.song@gsi.de}
\affiliation{GSI Helmholtzzentrum f\"{u}r Schwerionenforschung GmbH, Planckstrasse 1, 64291 Darmstadt, Germany}

\author{Elena Bratkovskaya}
\email{E.Bratkovskaya@gsi.de}
\affiliation{Helmholtz Research Academy Hesse for FAIR (HFHF), GSI Helmholtz Center for Heavy Ion Physics, Campus Frankfurt, 60438 Frankfurt, Germany}
\affiliation{Institut f\"{u}r Theoretische Physik, Johann Wolfgang Goethe-Universität,Max-von-Laue-Straße 1, D-60438 Frankfurt am Main, Germany}
\affiliation{GSI Helmholtzzentrum f\"{u}r Schwerionenforschung GmbH, Planckstrasse 1, 64291 Darmstadt, Germany}

\author{J\"org Aichelin}
\email{aichelin@subatech.in2p3.fr}
\affiliation{SUBATECH, Nantes University, IMT Atlantique, IN2P3/CNRS
4 rue Alfred Kastler, 44307 Nantes cedex 3, France}
\affiliation{ Frankfurt Institute for Advanced Studies,
  Ruth Moufang Str. 1, 60438 Frankfurt, Germany}

\date{\today}

\begin{abstract}
The two-particle correlation function measured in femtoscopic analyses provides access to the interaction potentials between emitted particles. This offers a unique opportunity to investigate interactions among charmed mesons and to explore the nature of possible exotic hadronic states. In this Letter, we study femtoscopic correlations of various charmed-meson pairs in relativistic heavy-ion collisions. The dynamical evolution of the system and charm hadron production are described within the Parton–Hadron–String Dynamics (PHSD) transport approach, while the correlation functions are computed using the Correlation Analysis Tool using the Schr\"odinger equation (CATS).
We demonstrate that heavy-ion collisions provide a significantly more favorable environment than $pp$ collisions for accessing charmed meson femtoscopic correlations. This arises from enhanced charm-quark production, reduced relative momenta due to in-medium energy loss, and a strong suppression of initial-state correlations. Our results indicate that femtoscopic measurements in heavy-ion collisions offer a sensitive probe of charmed meson interactions and possible hadronic molecular states.
\end{abstract}

\maketitle

\emph{Introudction.--}Quantum chromodynamics (QCD), the fundamental theory describing the strong interaction among quarks and gluons, allows for the existence of exotic hadronic states beyond the conventional quark model. These include glueballs composed solely of gluons~\cite{Mathieu:2008me,Ochs:2013gi}, hybrid states containing both quarks and gluons~\cite{Meyer:2015eta,Chanowitz:1982qj}, multiquark configurations such as tetraquarks and pentaquarks~\cite{Esposito:2016noz,Karliner:2017qhf}, as well as hadronic molecules~\cite{DeRujula:1976zlg,Guo:2017jvc}. In recent years, heavy-flavor exotic hadrons, particularly the hidden-charm states collectively referred to as the $\rm XYZ$ states, have attracted considerable attention and stimulated extensive theoretical and experimental investigations in the non-perturbative regime of QCD.

The first hidden-charm exotic hadron to be discovered was the $X(3872)$ state (also known as $\chi_{c1}(3872)$), observed by the Belle Collaboration in 2003~\cite{Belle:2003nnu}. Since its discovery, the properties and underlying structure of the $X(3872)$ have been intensively studied. The state has also been observed in heavy-ion collisions~\cite{CMS:2021znk}, providing an additional opportunity to probe its internal structure~\cite{ExHIC:2010gcb,Zhang:2020dwn,Wu:2020zbx,Chen:2021akx,Yun:2022evm}. Owing to its mass being extremely close to the $D^0\bar D^{*0}$ threshold, the $X(3872)$ has often been interpreted as a weakly bound meson--meson molecular state with a very small binding energy~\cite{Bignamini:2009sk}. Alternative interpretations include a compact tetraquark configuration, $c\bar c u\bar d$, or even a kinematic effect arising from mechanisms such as triangle singularities~\cite{Nakamura:2019nwd}. Despite substantial progress, the true nature of the $X(3872)$ remains an open question. To date, more than thirty $\rm XYZ$ states have been identified in experiments~\cite{Richard:2016eis,Hosaka:2016pey,Ali:2017jda,Liu:2019zoy}. 

The interaction potential between two heavy-flavor hadrons plays a crucial role in understanding possible molecular states and revealing their internal structure. One of the most effective experimental approaches for investigating hadronic interactions is femtoscopy, which is based on the Hanbury-Brown--Twiss (HBT) intensity interferometry technique~\cite{HBTc,Pratt:1984su,Lisa:2005dd}. Femtoscopy probes the space--time structure of the particle-emitting source at freeze-out~\cite{Lisa:2005dd,NA49:2007fqa,Li:2008qm,Wiedemann:1996ig,Kisiel:2006is} and establishes a connection between the experimentally measured two-particle correlation function and the underlying hadronic interaction~\cite{Lisa:2005dd}.

Benefiting from significant advances in experimental capabilities, numerous two-hadron correlation measurements have been performed in both proton--proton and heavy-ion collisions at the Relativistic Heavy Ion Collider (RHIC)~\cite{STAR:2005rpl,STAR:2018uho,STAR:2014dcy,STAR:2015kha} and the Large Hadron Collider (LHC)~\cite{ALICE:2018ysd,ALICE:2021cpv,ALICE:2019buq,ALICE:2019hdt,ALICE:2022uso,ALICE:2011kmy}. These measurements provide valuable opportunities to investigate hadronic interactions and possible exotic bound states. More recently, femtoscopic studies have been extended to the heavy-flavor sector. In particular, the ALICE Collaboration has measured the correlations of $pD^-$ and $\bar pD^+$ pairs in high-multiplicity $pp$ collisions at $\sqrt{s}=13~\mathrm{TeV}$~\cite{ALICE:2022enj}.

On the theoretical side, femtoscopy has recently been proposed as a promising tool for probing the internal structure of exotic hadrons such as the $X(3872)$~\cite{Kamiya:2022thy}, $X(3700)$~\cite{Abreu:2025jqy}, $Z_c(3900)$~\cite{Liu:2024nac}, $X(3960)$~\cite{Liu:2025wwx}, and $T_{cc}$~\cite{Albaladejo:2023wmv,Kamiya:2022thy,Ge:2026moy} in elementary collisions such as $pp$ or $e^+e^-$. However, experimental investigations remain challenging because of the relatively low production rate of charm quarks in these collision systems.

In this context, relativistic heavy-ion collisions offer several important advantages for studying heavy-flavor hadron interactions through femtoscopy. First, a ``charming'' medium characterized by an abundant production of charm quarks is created in heavy-ion collisions at LHC energies. As a result, the yield of charmed hadrons is significantly enhanced compared to that in $pp$ and $e^+e^-$ collisions. Second, femtoscopic correlations are most sensitive to hadron pairs with small relative momentum. Such conditions are naturally realized in heavy-ion collisions due to the substantial energy loss and medium-induced interactions experienced by heavy quarks while propagating through the quark--gluon plasma (QGP).
Therefore, heavy-ion collisions provide an ideal environment for investigating interactions between heavy-flavor hadrons and for exploring the nature of exotic hadrons from a femtoscopic perspective.

\emph{Space-time distribution of charmed mesons.--}
In our approach the dynamical evolution of the (anti)charm quarks after production is described by PHSD~\footnote{The website: https://phsd-phqmd.github.io}, which is a microscopic transport approach based on the first order gradient expansion of the Kadanoff-Baym equations, taking into account the off-shellness of particles for their strong interactions~\cite{Cassing:2008sv,Cassing:2009vt,Bratkovskaya:2011wp,Moreau:2019vhw}. Based on the effective dynamical quasi-particle model (DQPM) for the description of the QGP phase, it has given a good description of the experimental data on open and hidden heavy flavors in relativistic heavy-ion collisions~\cite{Song:2015ykw,Song:2015sfa}. Due to the large mass, the charm quarks are produced in initially and can be calculated by the perturbative QCD (pQCD). In our study, the energy-momentum of (anti)charm quarks is provided by the PYTHIA event generator~\cite{Sjostrand:2006za,Song:2015ykw}, in which the transverse momentum and rapidity are tuned to match those in the Fixed-Order Next-to-Leading Logarithm (FONLL) calculations~\cite{Cacciari:2012ny}. In heavy-ion collisions, nuclear shadowing effects are also included and are implemented using the EPS09 parametrization~\cite{Eskola:2009uj}.

\begin{figure}[!htb]
{ \includegraphics[width=0.45\textwidth]{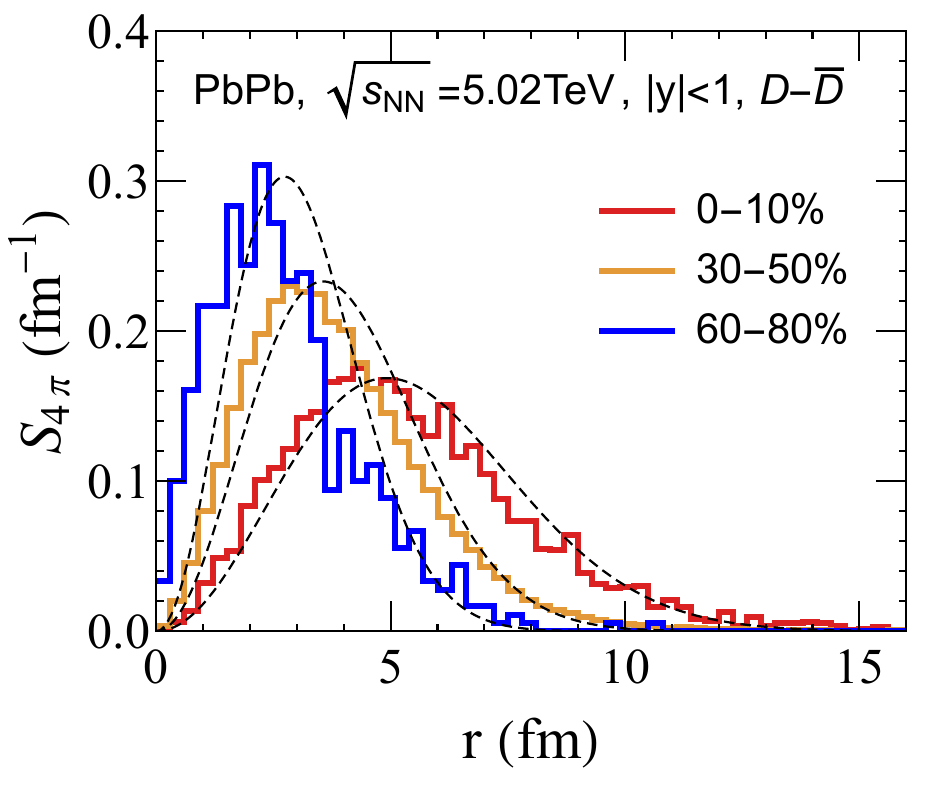}
\caption{The PHSD results for the probability density function of the relative distance $r$ of $D-\bar D$ pairs in their CoM frame produced in 0-10\% (red line), 30-50\% (orange line), and 60-80\% (blue line) PbPb collisions at $\sqrt{s_{\rm NN}}=5.02~\rm TeV$. The black dashed lines show the Gaussian sources with the same RMS radius.}
\label{fig.source.PbPb}}
\end{figure}

The charm quarks hadronize into various charmed hadrons through a combination of coalescence and fragmentation once the local energy density drops below $0.5~\rm GeV/fm^3$. Both the ground states ($D^{0,\pm}$ and $D_s$) and excited states ($D^{*0,\pm}$, $D_s^*$, $D^*_0(2300)^0$, $D_1(2420)^0$, and $D^*_2(2460)^{0,\pm}$) are included in the present study. More details can be found in Ref.~\cite{Song:2015sfa}.
The produced ground-state and excited charmed mesons subsequently scatter with light hadrons during the hadronic expansion. These interactions are described within heavy-quark effective theory~\cite{Abreu:2011ic,Song:2015sfa}. Once the inelastic scatterings cease and before the mesons decays the space-time information of the charmed mesons is recorded.

It should be noted that charmed mesons may freeze out at different times. If the time interval between the emission of two $D$ mesons is sufficiently large compared to the interaction timescale, or equivalently if their spatial separation exceeds the interaction range, the two $D$ mesons will no longer interact with each other. Consequently, in this study, we select hadrons with a time difference of $|t_1-t_2|<0.5~\rm fm/c$ in the nucleus-nucleus center-of-mass frame.  

The emission source, defined as the relative-distance distribution of $D-\bar D$ pairs (here $D$ means $D^0$ and $D^+$, $\bar D$ are their anti-particles) in their CoM frame, is calculated in heavy-ion collisions for the 0–10\%, 30–50\%, and 60–80\% centrality classes from PHSD and shown in Fig.~\ref{fig.source.PbPb}. It is compared with the commonly used spherical Gaussian parametrization (as shown with black dashed lines), $S_{4\pi}(r)=4\pi r^2/(2\sqrt{\pi}r_0)^3\exp (-{r^2/4r_0^2})$, 
where $r_0$ is the parameter that controls the size of the emission source. The same root-mean-square (RMS) radius gives $r_0=2.46,1.78,1.37~\rm fm$ for the 0-10\%, 30-50\%, 60-80\% centrality bins. 
We find the realistic emission source exhibits pronounced non-Gaussian features in peripheral collisions. In contrast, for central collisions the source is well approximated by a Gaussian form.
\begin{figure}[!htb]
{ \includegraphics[width=0.4\textwidth]{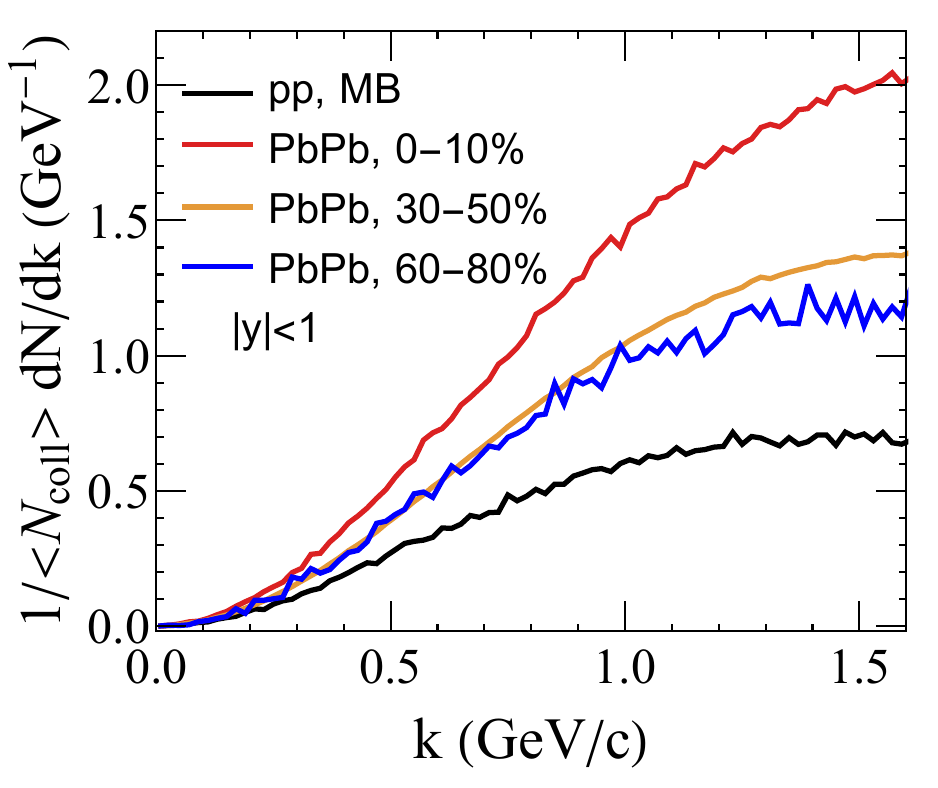}
\caption{The PHSD results for the distribution of the relative momentum $k$ of $D-\bar D$ pairs in their CoM frame produced in 0-10\% (red line), 30-50\% (orange line), 60-80\% (blue line) PbPb collisions, and minimum bias (MB) $pp$ collisions (black line) at $\sqrt{s_{\rm NN}}=5.02~\rm TeV$. The results are presented only in the low-k region, where the femtoscopic correlation is dominant.}
\label{fig.dk.PbPb}}
\end{figure}

As mentioned in the Introduction, a significantly larger number of charm quarks, and consequently $D$ mesons, are produced in heavy-ion collisions compared to $pp$ collisions. Moreover, due to the energy loss in the QGP created in heavy-ion collisions, which is shown clearly in nuclear modification factor $R_{AA}$, the relative momentum between $D$ and $\bar D$ mesons becomes smaller. In Fig.~\ref{fig.dk.PbPb}, we present the relative momentum $k$ distribution obtained from the PHSD. We can see that, after scaling by the number of collisions $N_{\rm coll}$, at low $k$ in each centrality class, the distribution in heavy-ion collisions is nearly twice as large as that in $pp$ collisions. This enhancement originates from correlations among different $D\bar D$ pairs, produced in the same heavy-ion event. They significantly increase the number of low-relative-momentum pairs and thus improve the available statistics. As a consequence, heavy-ion collisions provide a particularly favorable environment for studying $D\bar D$ femtoscopic correlations.

\emph{Hadronic interaction.--}The final-state interactions among various charmed hadrons are not incorporated in the PHSD model. The interaction potentials between two pseudoscalar charmed mesons, as well as between pseudoscalar and vector charmed mesons, have been investigated within the framework of effective field theories~\cite{Liu:2009qhy,Yun:2022evm}. These approaches are based on the principles of chiral symmetry and heavy quark symmetry, as discussed in Ref.~\cite{Liu:2009qhy}. A systematic study of possible molecular states composed of a pair of heavy mesons, such as $D\bar D$ and $D^*\bar D$, has been carried out in the meson-exchange model framework. The exchanged mesons include pseudoscalar and vector mesons, such as $\pi$, $K$, $\eta$, $\rho$, $\omega$, and $\phi$.

In order to account for the structure effect (internal structure and spatial distribution) of every interaction vertex, the monopole type form factor is introduced,
$F(q)=(\Lambda^2-m^2)/(\Lambda^2-q^2)$
where $q$ denotes the four-momentum of the exchanged meson. $\Lambda$ is a phenomenological parameter around 1 GeV.
The parameter $\Lambda$ governs the ``softness" of the interaction. A larger $\Lambda$ implies a ``harder" particle with less internal structure, while a smaller $\Lambda$ implies a ``softer" particle with more significant internal structure effects. This parameter is not fixed by the theory and will change the potential~\cite{Liu:2009qhy}. 
For the charged charmed mesons, the Coulomb interaction is needed to be considered as well.
In this study we average over the spin following \cite{Liu:2009qhy}.
In experiments, correlations between different charmed mesons are reconstructed from their decay products. However, this reconstruction procedure is not simulated in the present study. 

\emph{Femtoscopic Correlations.--} 
The $c$ and $\bar c$ quarks produced at the same production vertex are intrinsically correlated~\cite{Zhu:2006er,Zhao:2024oma}. These initial correlations are reflected not only in the angular distribution between the $c$ and $\bar c$ quarks, but also in their momentum distributions, which ultimately influence the measured femtoscopic observables. Therefore, in heavy-meson systems, the observed correlations are not generated solely by final-state interactions.

\begin{figure}[!hbt]
\centering
\includegraphics[width=0.4\textwidth]{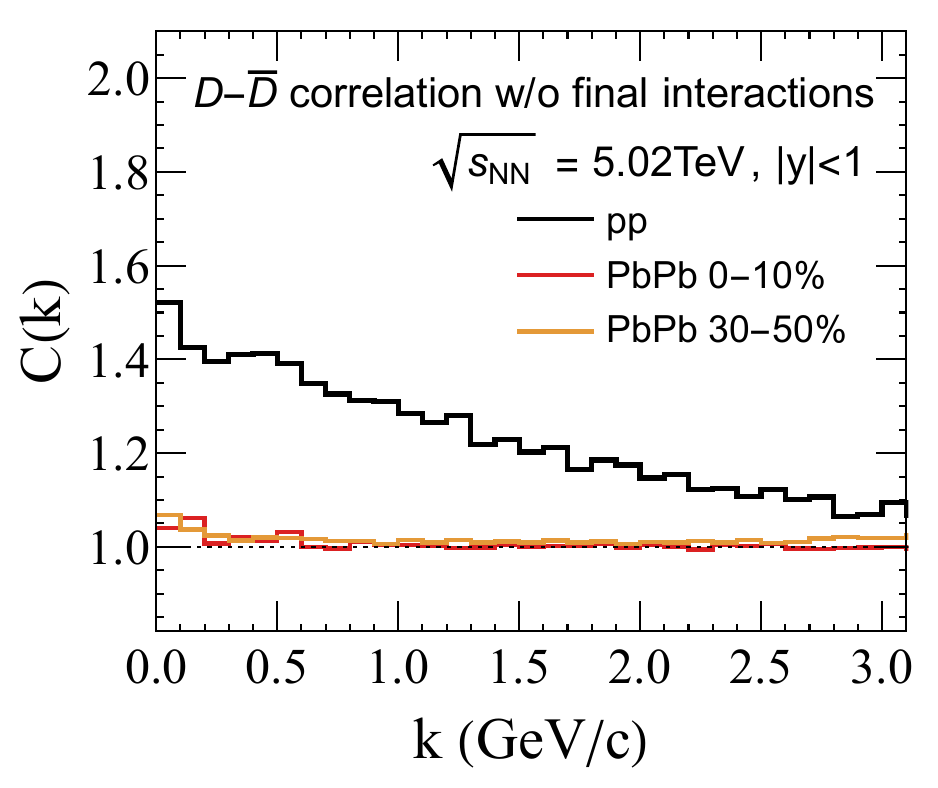}
\caption{Correlation functions of $D-\bar D$ in $pp$ (black line) and 0-10\% (red line), 30-50\% (orange line) PbPb collisions at $\sqrt{s_{\rm NN}}=5.02~\rm TeV$. All possible pairs of D mesons, produced in PHSD, are considered.}
\label{fig.corrwop}
\end{figure}

In $pp$ collisions, the relatively small number of produced $c\bar c$ pairs preserves these initial-state correlations. In contrast, in heavy-ion collisions a large number of charm--anticharm pairs are produced, leading to a substantial dilution of the initial correlations such that the emitted heavy mesons become effectively uncorrelated~\cite{Zhao:2024oma}. To illustrate this effect, we calculate the $D\bar D$ correlation function by definition,  
\begin{eqnarray}
C(k)= \mathcal{N}{N_{\rm same-pair}(k)\over N_{\rm mixed-pair}(k)},
\label{eq.corrmix}
\end{eqnarray}
where $N_{\rm same-pair}(k)$ and $N_{\rm mixed-pair}(k)$ are normalized relative momentum distribution of $D-\bar D$ pairs from the same and different events (mixed pairs), respectively. The normalization constant $\mathcal{N}$ is chosen to make the $C(k)\to 1$ at large $k$. This is commonly used in experiments. 
As mentioned above, the final-state interactions between $D-\bar D$ pairs are not incorporated in the PHSD model. Therefore, the observed correlation mainly originates from the initial stage and can serve as a baseline for assessing the effects of final-state interactions. The corresponding results are shown in Fig.~\ref{fig.corrwop}.

One observes that, for PbPb collisions, the correlation function remains very close to unity, indicating the absence of significant correlations. On the other hand, in $pp$ collisions the mesons exhibit sizable correlations even at large relative momenta. This behavior differs qualitatively from correlations induced by final-state interactions, which are expected to predominantly affect the low-relative-momentum region.

Now, we study the influence of the final state interaction between the $D$-mesons on the correlation function. The correlations can be calculated using the Koonin–Pratt formalism~\cite{Koonin:1977fh,Pratt:1986cc}. In this approach, the correlation function is obtained by convoluting the source function $S(\bm r)$ with the two-body scattering wave function, $\psi_k(\bm r)$,
\begin{eqnarray}
C(k)=\int S({\bm r})|\psi_k({\bm r})|^2d{\bm r},
\label{eq.correlation}
\end{eqnarray}
where $k=|{\bf p}_1^*-{\bf p}_2^*|/2$ denotes the relative momentum of the particle pair in their center-of-mass frame, and ${\bm r}$ is the relative distance between the two particles. $m_1$ and $m_2$ represent the masses of the two hadrons. The scattering wave function $\psi_k({\bm r})$ is obtained by solving the radial Schr\"odinger equation,
\begin{eqnarray}
\frac{d^2 u_{k,l}(r)}{dr^2} = \left( 2\mu V_i(r) + \frac{l(l+1)}{r^2} - k^2 \right) u_{k,l}(r),
\label{eq.schroedingeq}
\end{eqnarray}
where $\mu = m_{D_1}m_{D_2}/(m_{D_1}+m_{D_2})$ is the reduced mass of two charmed meson states, and $u_{k,l}(r)$ represents the radial wave function in $l-$wave scattering. 
$V_i$ is taken from the aforementioned interaction potentials for each pair, with $\Lambda=0.5~\rm GeV$. For low-energy scattering, the $S-$wave channel dominates. However, as the energy increases, contributions from higher-$l$ channels become significant. Consequently, the total scattering wavefunction can be expressed as follows, $\psi_k({\bf r})=\sum_{l=0}^{l_{\rm max}}(2l+1)i^l(u_{k,l}/r)P_{l}(\cos \theta)$, where $P_l$ denotes the Legendre polynomials. 
For low-energy scattering, the series converges relatively rapidly. In this study, we utilized the CATS to solve the Schr\"odinger equation~\cite{Mihaylov:2018rva,Fabbietti:2020bfg}.
\begin{figure}[!hbt]
\centering
\includegraphics[width=0.24\textwidth]{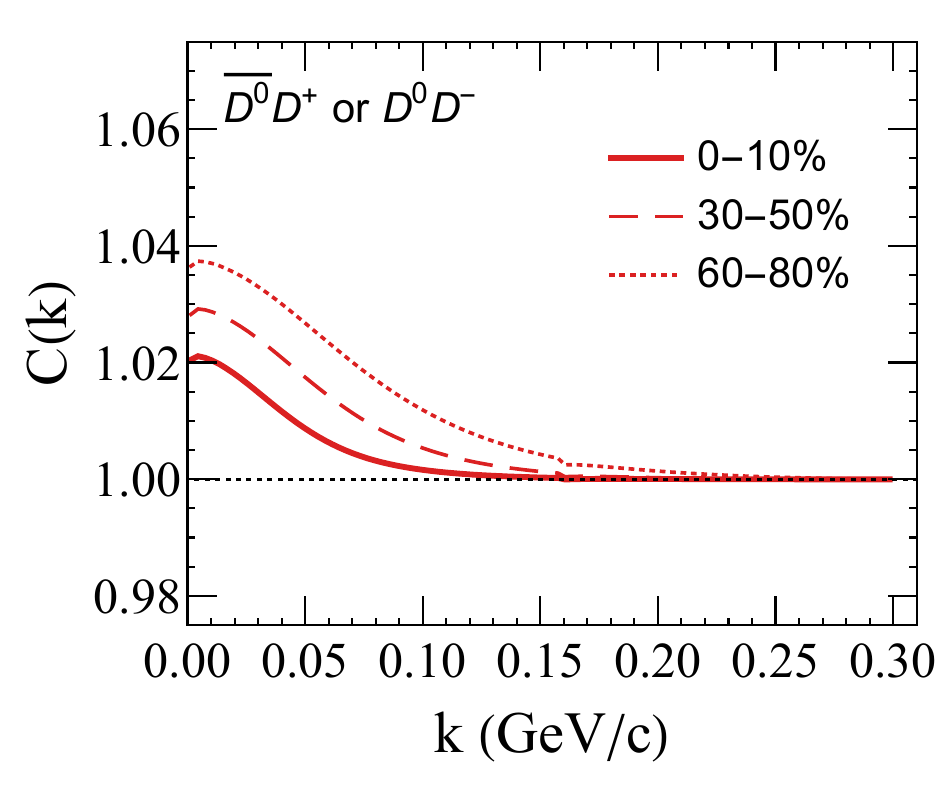}\includegraphics[width=0.24\textwidth]{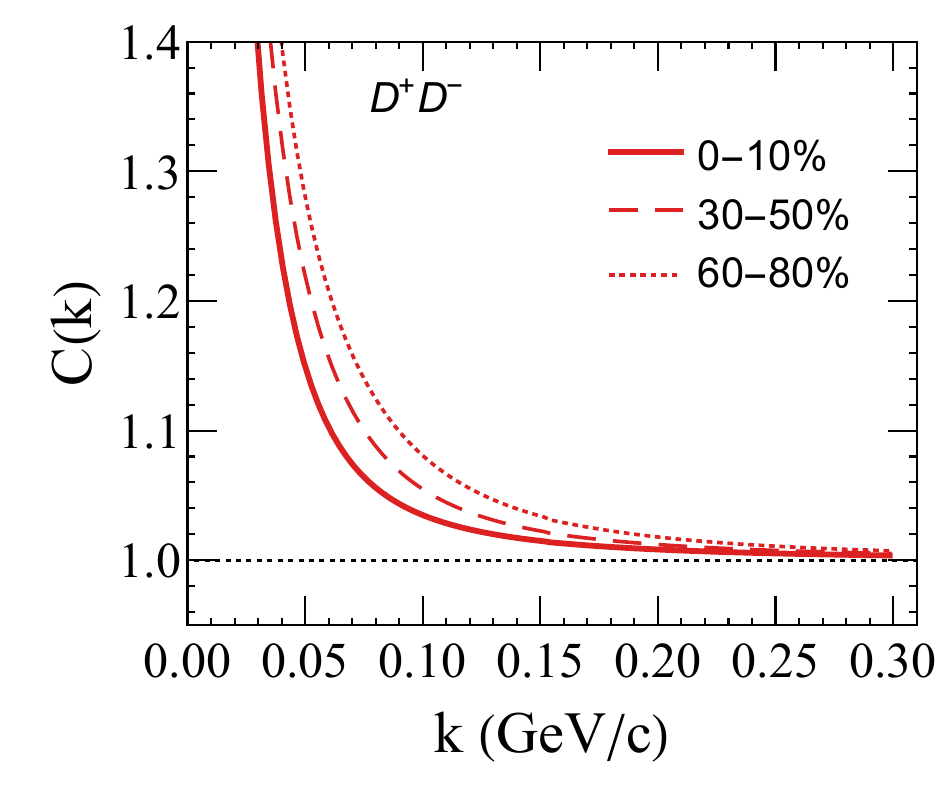}\\
\includegraphics[width=0.24\textwidth]{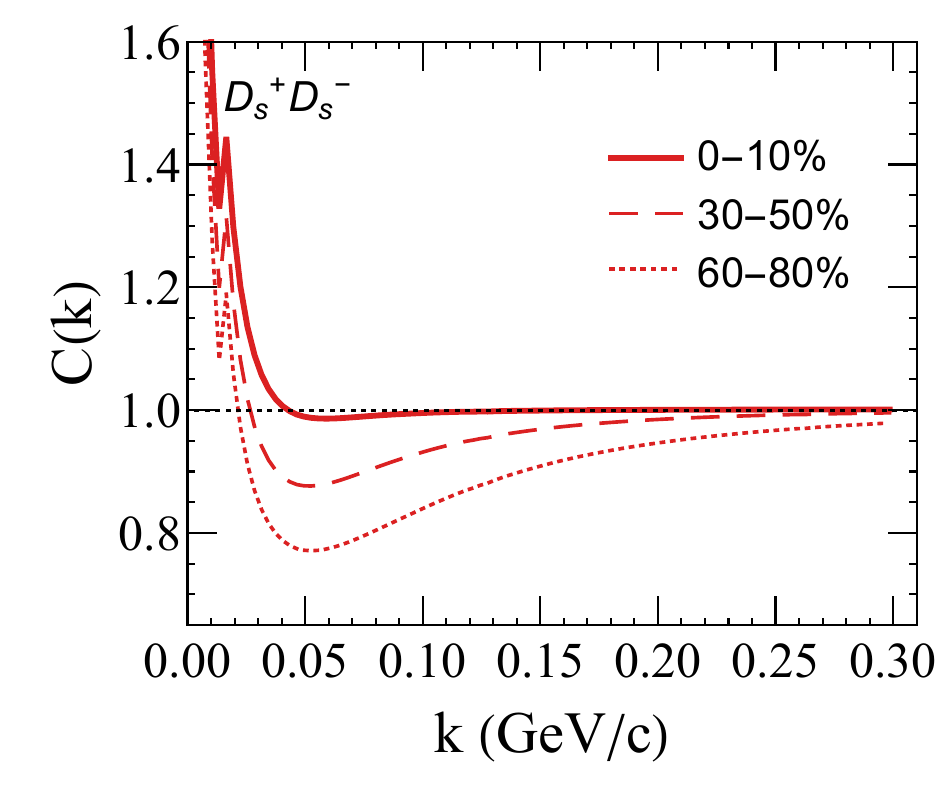}\includegraphics[width=0.24\textwidth]{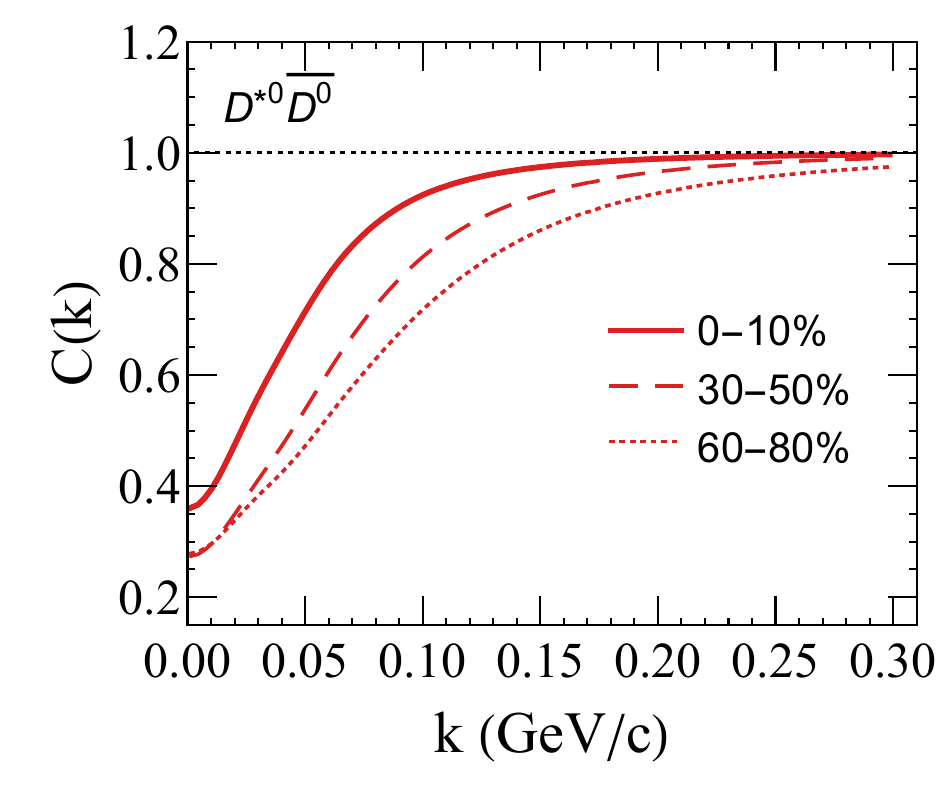}
\caption{The correlation function of $\bar D^0-D^+$, $D^+-D^-$, $D_s^+-D_s^-$, and $D^{*0}-\bar D^0$, in 0-10\% (solid line), 30-50\% (dashed line), and 60-80\% (dotted line) PbPb collisions with $|y|<1$ at $\sqrt{s_{\rm NN}}=5.02~\rm TeV$.}
\label{fig.corrAA}
\end{figure}

\begin{figure}[!hbt]
\centering
\includegraphics[width=0.4\textwidth]{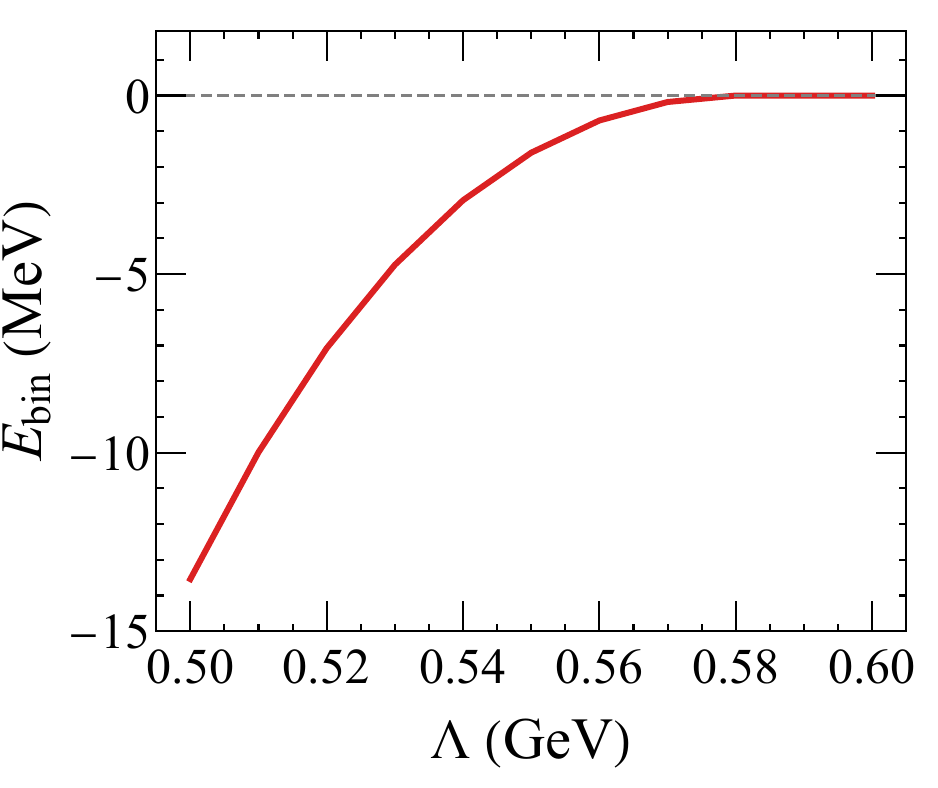}\\
\includegraphics[width=0.4\textwidth]{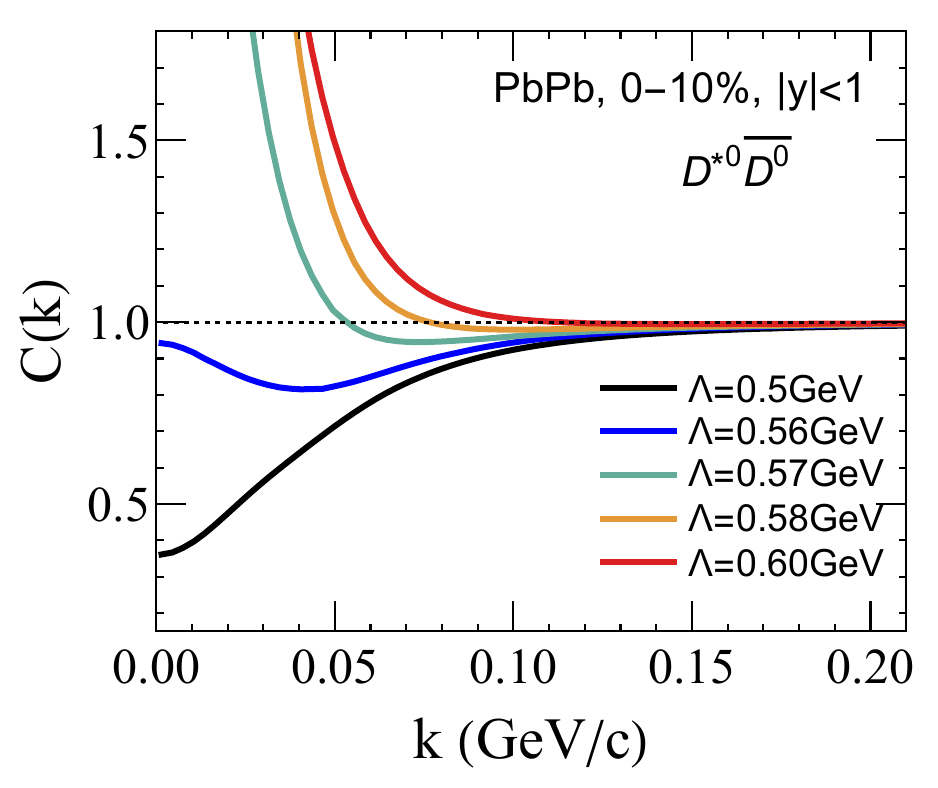}
\caption{The binding energy $E_{\rm bin}$ and correlation function of $D^{*0}-\bar D^0$ in 0-10\% PbPb collisions at $\sqrt{s_{\rm NN}}=5.02~\rm TeV$. The correlations are calculated with several $\Lambda$s (lower panel).}
\label{fig.corrAA2}
\end{figure}

Knowing the scattering wavefunctions and emission source functions, we can use Eq.~\eqref{eq.correlation} to calculate the correlation functions between various charmed hadrons. The correlation functions for different centrality bins are shown in Fig.~\ref{fig.corrAA}. We can see that, due to a very short and weak interaction between $D^0$ and $D^{\pm}$, their correlation is very small and close to 1, which is hard to observe in experiments. For the $D^+-D^-$ system, a very strong positive correlation is observed at a very small relative momentum $k$, driven by the attractive Coulomb interaction. In general, an attractive interaction tends to enhance the femtoscopic correlation function. However, when a bound or quasi-bound state is formed, the correlation function at low $k$ can be smaller than 1, as seen in the $D_s^+-D_s^-$ system. For the $D^{*0}-\bar D^0$ system, the absence of the Coulomb interaction implies that the correlation function is governed solely by the strong hadronic interaction and remains negative over the entire relative-momentum range. For all systems, the correlation function becomes stronger from central to peripheral collisions, reflecting the decreasing size of the emission source.

We now examine the sensitivity of the correlation function to the interaction strength. Focusing on the $D^{*0}-\bar D^0$ system in 0–10\% Pb–Pb collisions, we vary the cutoff parameter $\Lambda$ from 0.5 to 0.6 GeV. For each value, we solve both the static and scattering Schr\"odinger equations to determine the bound-state properties and the corresponding correlation functions. The results are shown in Fig.~\ref{fig.corrAA2}. When the $\Lambda>0.57~\rm GeV$, the bound state disappears. This transition is clearly reflected in the correlation function, which becomes entirely positive. 
Therefore, the correlation function provides a clear experimental signature for the existence of a $D^{*0}-\bar D^0$ molecular state. 
If the system is tightly bound, the correlation function becomes entirely negative. For a loosely bound state, the correlation function exhibits a shallow dip at intermediate relative momentum, followed by a strong enhancement at very low relative momentum. In the absence of a bound state, the correlation function remains positive over the full momentum range.

\emph{Summary.--}
In this Letter, we investigated femtoscopic correlations between various charmed meson pairs in relativistic heavy-ion collisions. The dynamical evolution and hadronization of charm quarks were described within the PHSD transport approach. After freeze-out, the interactions among the produced charmed mesons were incorporated through interaction potentials derived from effective field theory and light-hadron exchange models. The resulting correlation functions were calculated using the CATS framework.

Compared to $pp$ collisions, heavy-ion collisions provide a particularly favorable environment for studying heavy-flavor meson correlations for several reasons: (i) the abundant production of heavy quarks and heavy-flavor mesons, (ii) the substantial in-medium energy loss of heavy quarks, which enhances the probability of producing hadron pairs with small relative momentum, and (iii) the strong suppression of initial-state correlations, making the measured femtoscopic correlations predominantly sensitive to final-state interactions.

Our results show that the femtoscopic correlation function is highly sensitive to the underlying interaction potential between heavy mesons. In particular, using the $D^{*0}\bar D^0$ system as an example, we demonstrate that femtoscopy can serve as a powerful probe of possible molecular bound states and thus provide valuable information on the internal structure of exotic hadrons. These measurements are expected to become accessible with the high-statistics data from the upgraded LHC experiments.

~\\
\text{Acknowledgment.} 
We acknowledge the support by the Deutsche Forschungsgemeinschaft (DFG, German Research Foundation) through the grant CRC-TR 211 'Strong-interaction matter under extreme conditions' - Project number 315477589 - TRR 211. The computational resources have been provided by the Center for Scientific Computing (CSC) of the Goethe University, Frankfurt.

\bibliographystyle{apsrev4-1}
\bibliography{refs}

\end{document}